\documentclass[aps,pra,reprint,onecolumn,notitlepage,eqsecnum,floatfix,showkeys,nofootinbib]{revtex4-2}

\usepackage{amsmath}
\usepackage{amsbsy}
\usepackage{amsfonts}
\usepackage{bigints}
\usepackage{xcolor}
\usepackage{graphicx}
\usepackage{hyperref}
\usepackage{multirow}
\usepackage{linearb}
\usepackage{relsize}
\usepackage{amsthm}
\usepackage{fancyvrb}
\usepackage{calrsfs}
\DeclareMathAlphabet\mathbfcal{OMS}{cmsy}{b}{n}

\newcommand{\bq}{\begin{eqnarray}}
\newcommand{\eq}{\end{eqnarray}}
\newcommand{\bqn}{\begin{eqnarray*}}
\newcommand{\eqn}{\end{eqnarray*}}
\newcommand{\bqs}{\begin{subequations}}
\newcommand{\eqs}{\end{subequations}}
\newcommand{\bw}{\begin{widetext}}
\newcommand{\ew}{\end{widetext}}
\newcommand{\xx}{{\boldsymbol x}}

\newcommand{\rr}{{\boldsymbol r}}

\newcommand{\qq}{{\boldsymbol q}}

\newcommand{\calm}{{\cal M}}
\newcommand{\calo}{{\cal O}}

\newcommand{\cals}{{\cal S}}

\newcommand{\cald}{{\cal D}}

\newcommand{\calp}{{\cal P}}
\newcommand{\calr}{{\cal R}}

\newcommand{\calh}{{\cal H}}
\newcommand{\calc}{{\cal C}}
\newcommand{\cala}{{\cal A}}

\newcommand{\calx}{{\cal X}}
\newcommand{\caly}{{\cal Y}}

\newcommand{\calu}{{\cal U}}

\newcommand{\calk}{{\cal K}}

\newcommand{\id}{1\!\!1}

\newcommand{\red}[1]{{#1}}

\SaveVerb{floor}=FLOOR=
\SaveVerb{int}=INT=
\SaveVerb{nint}=NINT=
\SaveVerb{sgn}=SGN=
\SaveVerb{abs}=ABS=

\begin{document}
\title{Path Integral Monte Carlo on a Sphere}

\author{Riccardo Fantoni}
\email{riccardo.fantoni@scuola.istruzione.it}
\affiliation{Universit\`a di Trieste, Dipartimento di Fisica, strada
  Costiera 11, 34151 Grignano (Trieste), Italy}

\date{\today}

\begin{abstract}
We solve numerically exactly a simple toy model to quantum general relativity or 
more properly to path integral on a curved space. We consider the thermal
equilibrium of a quantum many body problem on the sphere, the surface of 
constant positive curvature. We use path integral Monte Carlo to measure
the kinetic energy, the internal energy and the static structure of a 
bosons, fermions and anyons fluid at low temperatures on the sphere. For 
bosons we also measure the superfluid fraction and compare its behavior 
at the critical temperature with the universal jump predicted by Nelson and 
Kosterlitz in flat space in the thermodynamic limit at the superfluid phase 
transition. For fermions and anyons it 
is necessary to use the restricted path integral recipe in order to overcome 
the sign problem. Even if this recipe is exact for the non interacting fluid it
reduces to just an approximation for an interacting system. And we make the 
example of the electron gas at low temperature. Snapshots of the many body
path configuration during the evolution of the computer experiment show 
that the ``speed'' of the single particle path near the poles slows down
as a consequence of the ``hairy ball theorem'' of Poincar\'e. The influence 
of curvature on the thermodynamic and structural properties of the many
body fluid is also studied.
\end{abstract}


\maketitle
\tableofcontents
\section{Introduction}
\label{sec:intro}

One of the greatest challenges of today physics is the problem of putting 
together the quantum theory with the theory of general relativity, or from the
mathematical side we are looking forward to a bridge between the functional
integral underlying the time evolution in a Hilbert space and the Riemannian 
manifold underlying the support of
differential geometry or more generally differential topology. We are still
at the beginning of this ambitious project and from this point of view any
simple, yet well defined, physical model at the intersection of these two 
theories is certainly valuable, especially so if the model can be solved 
exactly from a mathematical point of view \cite{Fantoni23b,FantoniCQ}.

From the laboratory experimental point of view we must stress that in 
recent years some extraordinary advances have been made in the manufacturing
of cold quantum gases trapped on various geometries 
\cite{Tononi2023,Tononi2024a}. The advances in the realization of 
trapped {\sl Bose-Einstein condensates} 
\cite{Tononi2019,Tononi2020,Tononi2022a,Tononi2022b,Cinti2024,Tononi2025,Cinti2025} 
has interests in the construction of quantum interferometry sensors, 
atomic lasers, or quantum computers. Whereas trapped electron liquids 
have been realized in nuclear fusion experiments. Only recently we have 
been able to create a white dwarf artificially in a earthly laboratory 
\cite{wikifusion}. This is one more reason to study the properties of 
{\sl Jellium} \cite{Kenny1996,Brown2013,Fantoni21b,Fantoni21i}, an 
electron liquid where the ionic component is approximated by a uniform 
neutralizing background stabilizing the plasma. This 
is just a first brute approximation to the more realistic model of a two 
component plasma \cite{Ceperley1981,Ceperley1987}; 
but even so it poses the extremely challenging problem of the determination 
of the ground state and low temperature properties of a many electron 
system on a curved spacetime \cite{Fantoni18c,Fantoni23a} with the additional 
subtleties of overcoming the fermions sign problem 
\cite{Ceperley1980,Ceperley1991,Ceperley1995} and ordering problems on 
properly defined self adjoint operators subject to holonomic 
constraints as the ones faced in the development of a quantum theory of 
curved spacetime \cite{Fantoni23b,Fantoni24f,Fantoni25a,Fantoni25g}.  

Guided by these motivations some years ago we studied the statistical physics 
problem of an electron gas at finite, non zero temperature on the surface of
a sphere \cite{Fantoni18c,Fantoni23a}, {\sl the} surface of constant positive 
curvature and probably the simplest of all Riemannian manifolds. Similar
studies on the Haldane sphere had already been tried but at zero temperature
\cite{Ortiz1997a,Ortiz1997b,Ortiz2000}. And at high temperature, at a special 
value of the Coulomb coupling constant, in the non quantum regime, on several 
different (curved) surfaces: the plane  
\cite{Ginibre65,Metha67,Jancovici81b,Alastuey81}, the cylinder 
\cite{Choquard81,Choquard83}, the sphere  
\cite{Caillol81,Tellez99,Jancovici00,Salazar2016}, the pseudosphere 
\cite{Jancovici1998,Fantoni03a,Jancovici04}, and the Flamm paraboloid 
\cite{Fantoni08c, Fantoni12b}. \red{Other authors \cite{Tempere2002} studied
the electron gas on a sphere within linear response theory}.

Here we want to examine carefully the case of a many body system 
on a sphere at finite, non zero low temperature, i.e. in its quantum regime.
The thermal equilibrium properties of the many body system
will be captured by a path integral description of its 
density matrix on the sphere.

The bodies that form the statistical physics model can be described either
as distinguishable or as identical. In their description as identical bodies 
we can further distinguish between whether they obey to Bose-Einstein
statistics or if they obey to Fermi-Dirac statistics according to their
wave function transformation property under their permutation. If additionally 
they are described as identical and impenetrable bodies they will obey to 
anyonic statistics according to their wave function imaginary time braiding
evolution. The braid group was introduced in 1925 by Emil Artin
\cite{Artin1947}.  
All these descriptions bring to different statistical physics and 
thermodynamic properties when the many body system is in its low temperature
quantum regime, and they merge at high temperature \cite{Fantoni12c}.  

On the other side, already the simple sphere has some quite delicate 
features as, for example, the {\sl hairy ball theorem}, according to which 
her Euler class is the obstruction to her tangent planes, the {\sl tangent 
bundle},
\footnote{A particular {\sl fiber bundle}.}
 having always a non vanishing {\sl fiber}, or hair, for any 
{\sl section}
\footnote{In topology, a cross {\sl section} of a fiber (tangent) 
bundle space, $B\times F$ is a graph over 
the {\sl base space} $B$, in this case the sphere. 
A choice of a tangent vector to any point of the sphere 
is a section of the tangent bundle of the sphere.}.
The theorem was first proven by Henri Poincar\'e for 
the sphere in 1885 \cite{Poincare1885}, and extended to higher even 
dimensions in 1912 by Luitzen Egbertus Jan Brouwer \cite{Brouwer1912}.
The theorem has been expressed colloquially as ``you can't comb a 
hairy ball flat without creating a cowlick''.
If $f$ is a continuous function that assigns a vector in the three 
dimensional space to every point $\rr$ on a sphere such that $f(\rr)$ 
is always tangent to the sphere at $\rr$, then there is at least one 
{\sl pole}, a point where the field vanishes, i.e. an $\rr$ such that
$f(\rr)=0$. Every zero of a vector field has a (non-zero) {\sl index}
\footnote{The index of a bilinear function/al is the dimension of the
space on which it is negative definite. In the context of vector fields 
on a Riemannian manifold the index is equal to $+1$ around a source or 
a sink, and more generally equal to $(-1)^k$ around a saddle that has 
$k$ contracting dimensions and $n-k$ expanding dimensions.}, 
and it can be shown that the sum of all of the indexes at all of the 
zeros must be two, because the Euler characteristic of the sphere is two. 
Therefore, there must be at least one zero. This is a consequence of the 
{\sl Poincar\'e-Hopf theorem}. The theorem was proven for two dimensions 
by Henri Poincar\'e and later generalized to higher dimensions by Heinz 
Hopf \cite{Hopf1926}. In particular we expect that even a single free 
particle in thermal equilibrium at low temperature on the sphere have 
a path which will be subject to some peculiar topological features
\cite{Bastianelli2017}.

The question of the influence on the statistical physics properties of
a quantum many body fluid of the curvature of the supporting Riemannian 
surface is not straightforward. In fact already for a single free particle 
the phase space dynamics on a sphere is generally not ergodic since her
trajectories, the geodesics, may be confined to invariant tori, but the
picture dramatically changes on a pseudosphere, {\sl the} surface of constant 
negative curvature, as was proven by Emil Artin in 1924 \cite{Artin1924}, 
where geodesics diverge \cite{Fantoni03a,Fantoni19a}. 
This makes the quantum version of this system a paradigmatic model of 
quantum chaos.

We then here plan to study this preliminary simple toy model of a quantum
many body fluid on a sphere, which nonetheless does not allow for an analytic 
exact solution, with the instrument of 
Path Integral Monte Carlo (PIMC) \cite{Ceperley1995} which is able to 
extract exact numerical properties of the statistical physics model.
It certainly is gratifying to know that at least this simple toy model
of statistical general relativity \cite{Fantoni24f,Fantoni25a,Fantoni25g} 
can be solved exactly at least numerically.

\section{Many body path integral on a Riemannian manifold}
\label{sec:pirm}

Throughout the whole paper we will denote with a bold face letter a point on 
the $d$ dimensional Riemannian manifold $\calm$. Greek indexes run over the $d$ 
space dimensions. Roman indexes are used either for a particle label or for
an imaginary timeslice label in the path integral discretization. We use 
Einstein summation convention of tacitly assuming a sum over repeated greek 
indexes.

A many body system is composed of $N$ {\sl distinguishable} particles of mass 
$m$ and spin $s$ with positions in 
$R=(\rr_1,\rr_2,\ldots,\rr_N)=(\{\rr_i\})$ where each position vector
$\rr_i=(r_i^1,r_i^2,\ldots,r_i^d)=(\{r_i^\alpha\})$ in $d$ dimensions. 
On a Riemannian manifold $\calm$ of dimension $d$ and metric tensor 
$g_{\alpha\beta}(\rr)$,
the geodesic distance between two infinitesimally close points $R$ and $R'$ is
$d\tilde{s}^2(R,R')=\sum_{i=1}^Nds^2(\rr_i,\rr_i')$ where 
$ds^2(\rr,\rr')=g_{\alpha\beta}(\rr)(\rr-\rr')^\alpha(\rr-\rr')^\beta$. 
Moreover,  
\bq
\tilde{g}_{\mu\nu}(R)&=&
g_{\alpha_1\beta_1}(\rr_1)\otimes\ldots
\otimes g_{\alpha_N\beta_N}(\rr_N),\\
\tilde{g}(R)&=&\prod_{i=1}^N\det||g_{\alpha_i\beta_i}(\rr_i)||,
\eq
where $||\tilde{g}_{\mu\nu}||$ is a matrix made of $N$ diagonal blocks
$||g_{\alpha_i\beta_i}||$ with $i=1,2,\ldots,N$.
The Laplace-Beltrami operator on the manifold of dimension $dN$ is
\bq
\Delta_R=\tilde{g}^{-1/2}\nabla_\mu(\tilde{g}^{1/2}
\tilde{g}^{\mu\nu}\nabla_\nu),
\eq
where $\nabla=\nabla_R$, $\tilde{g}^{\gamma\nu}$ is the inverse of
$\tilde{g}_{\gamma\nu}$, i.e. 
$\tilde{g}_{\mu\gamma}\tilde{g}^{\gamma\nu}=\delta_\mu{}^\nu$ 
the Kronecker delta.

We will first assume {\sl free}, non interacting bodies, with an Hamiltonian 
$\calh$ that reduces to the one of the free gas in flat space. For the
sake of simplicity
\footnote{This is a delicate point especially for what concerns ordering 
ambiguities \cite{Fantoni23b}. We here appeal to simplicity.} we will choose
\bq \label{eq:Hcurved}
\calh=-\lambda\Delta_R,
\eq
with $\lambda=\hbar^2/2m$. 

For {\sl interacting} bodies we will then have more generally
\bq
\calh=-\lambda\Delta_R+V(R),
\eq
where $V(R)$ is the {\sl physical potential} energy of the system, that we 
here assume only a function of the particles positions and bounded from below.

The density matrix $\rho$ of the system obeys to Bloch equation
\bq
\frac{\partial\rho(t)}{\partial t}&=&-\calh\rho(t),\\
\rho(0)&=&\id,
\eq
where $t$ is the imaginary time with dimension of the reciprocal of
energy and $\id$ the identity matrix. 
The position representation of the density matrix is then obtained
from $\rho(R,R';t)=\langle R|\rho(t)|R'\rangle$ with 
$\langle R|R'\rangle=\delta(R-R')/\sqrt{\tilde{g}(R)}$ where 
$\delta$ is a $dN$ dimensional Dirac delta function.
In the small imaginary time $\tau$ limit 
the position representation of the density matrix is 
\bq \label{eq:rhocurved}
\rho(R,R';\tau)\propto
\tilde{g}(R)^{-1/4}\sqrt{\cald(R,R';\tau)}\,
\tilde{g}(R')^{-1/4}
e^{\lambda\tau\calr(R)/6}
e^{-\cals(R,R';\tau)},
\eq
where $\calr$ is the scalar curvature of the manifold
\footnote{The factor depending on the curvature of the manifold is due to 
Bryce DeWitt \cite{DeWitt1957}. 
For a space of constant curvature there is clearly no effect, 
as the term due to the curvature just leads to a constant multiplicative 
factor that has no influence on the measure of the various observables.},
$\cals$ the action, and $\cald$ the van Vleck's determinant 
\cite{Vleck1928,Schulman} 
\bq
\cald_{\mu\nu}&=&-\nabla_\mu\nabla'_\nu\cals(R,R';\tau),\\
\det||\cald_{\mu\nu}||&=&\cald(R,R';\tau),
\eq
where $\nabla=\nabla_R$ and $\nabla'=\nabla_{R'}$. This determinant is 
the Jacobian of the transformation from the initial conditions given by
fixing the pair of momentum and coordinate to the boundary conditions given by
specifying the pair of initial and final coordinates needed in the 
coordinate representation. The square root of the 
van Vleck determinant factor takes into account the density of paths among
the minimum extremal region for the action (see Chapter 12 of Ref. 
\cite{Schulman}). For the density matrix (\ref{eq:rhocurved}) 
the volume element of integration is $\sqrt{\tilde{g}(R)}\,dR$. 
The two factors $\tilde{g}^{-1/4}$ are needed 
in order to have for the density matrix a bidensity for which the boundary 
condition to Bloch equation is simply a Dirac delta function. In the 
small $\tau\to 0$ limit
\footnote{Remember that the metric tensor is covariantly constant.}
$\tilde{g}(R)^{-1/4}\sqrt{D(R,R';\tau)}\,\tilde{g}(R')^{-1/4}\to(1/2\lambda\tau)^N$.

For the {\sl action} $\cals$, the {\sl kinetic-action} $\calk$, and the 
{\sl inter-action} $\calu$ we have 
\footnote{The expression for $\calk$ is the one of Eq. (24.16) of Ref. 
\cite{Schulman} to lowest order in $R-R'$.}
\bq \label{eq:action}
\cals(R,R';\tau)&=&\calk(R,R';\tau)+\calu(R,R';\tau),\\ \label{eq:kineticaction}
\calk(R,R';\tau)&=&\frac{dN}{2}\ln(4\pi\lambda\tau)+
\frac{d\tilde{s}^2(R,R')}{4\lambda\tau}.
\eq
In particular the kinetic-action is responsible for a diffusion of the
random walk with a single particle variance
equal to $\sigma^2_{\alpha\beta}(\rr)=2\lambda\tau/g_{\alpha\beta}(\rr)$.
The inter-action is defined as $\calu=\cals-\calk$ and for potential
energies bounded from below one can resort to the Trotter formula 
\cite{Trotter1959} to reach the {\sl primitive approximation} 
\footnote{See Ref. \cite{Ceperley1995} for a numerical analysis of the
accuracy of this approximation and for its possible refinements.}
\bq
\calu(R,R';\tau)=\tau [V(R)+V(R')]/2.
\eq
For non interacting bodies $\calu=0$. Note that, even to lowest order in 
$R-R'$ \footnote{For next orders corrections see for example 
Ref. \cite{Bastianelli2017}.}, the path integral in the curved manifold
for the non interacting system will not coincide with the one in flat space
since it is not possible with a change of coordinates to simply remove 
the metric factor from both $d\tilde{s}^2$ and the volume element of 
integration, if not only locally. In fact this would require a {\sl non
coordinate basis} \cite{Gravitation}.  

Given then an observable $\calo$ we can determine its thermal average
at an absolute temperature $T$ from
\bq \label{eq:measureO}
\langle\calo\rangle&=&g_s{\rm tr}\{\rho(\beta)\calo\}/Z_N,\\
Z_N&=&g_s{\rm tr}\{\rho(\beta)\},
\eq
where $\beta=1/k_B T$ with $k_B$ Boltzmann constant, ${\rm tr}\{\ldots\}$ is 
the trace over the spatial variables, $Z_N$ the canonical partition function, 
$g_s=2s+1$ is the spin degeneracy, and we assumed the Hamiltonian independent 
from spin. For {\sl identical} bodies the spin-statistics 
theorem of quantum field theory, dictates that in spatial dimension bigger 
than two, particles with integer spin are bosons (obeying Bose-Einstein 
statistics, symmetric wavefunctions), while half-integer spin particles are 
fermions (obeying Fermi-Dirac statistics, antisymmetric wavefunctions). In 
dimension two anyonic statistics are also possible for identical and
{\sl impenatrable} bodies \cite{Lerda}. In this work, for bosons we will
just consider the {\sl spinless} $s=0$ case, for fermions the spin $s=1/2$
{\sl fully polarized} case, with only one spin species, either all
bodies spin up or all bodies spin down, and for anyons we make no 
assumption on their spin. In any case the measure (\ref{eq:measureO}) 
of the various thermodynamic or structural observables will not depend on 
$g_s$. 

The position representation of the density matrix at an imaginary 
time $t=\beta$ is obtained through a {\sl path integral}
\bq \label{eq:pi}
\rho(R,R';\beta)=\langle R|\rho(\beta)|R'\rangle
=\int\prod_{k=1}^{M-1}\left[\rho(R_k,R_{k+1};\tau)\,
\sqrt{\tilde{g}(R_k)}\,dR_k\right]
\delta(R_0-R)\delta(R_M-R')\,dR_0dR_M,
\eq
where we have discretized the imaginary time $\beta$ into $M$ 
{\sl timeslices} with a small {\sl timestep} $\tau=\beta/M$. 
The volume element of integration, $\sqrt{\tilde{g}(R)}\,dR$, can be
treated by introducing $W(R)=-\ln[\sqrt{\tilde{g}(R)}]/\tau$ as an 
effective {\sl geometrical potential}. The many body path
$R(t)$ is consequently discretized in $M$ {\sl beads} 
$R_k=(\{\rr_{i,k}\})=(\{r_{i,k}^\alpha\})$ at each 
timeslice $k=1,2,\ldots,M$. We will also call {\sl link} a pair of contiguous 
beads. Note that in order to measure an observable through Eq. (\ref{eq:measureO})
it is necessary to consider closed paths such that $R(t+\beta)=R(t)$, or rings 
on the manifold $\calm$.

For identical bodies, if they satisfy to the Bose-Einstein statistics
one needs to symmetrize the distinguishable density matrix, if they satisfy
to the Fermi-Dirac statistics one needs to antisymmetrize it \cite{Ceperley1996}. 
In these cases we can then write
\footnote{One can symmetrize or antisymmetrize respect to the first, the second
or both the arguments of the distinguishable density matrix. We here choose the
first case.}
\bq \label{eq:dmid}
\rho_\pm(R,R';\beta)&=&\frac{1}{N!}\sum_\calp{\rm sgn}(\calp)\rho(\calp R,R';\beta),\\
{\rm sgn}(\calp)&=&(\pm 1)^{\sum_{\nu=1}^N(\nu-1)C_\nu},
\eq
where $\calp$ is any permutation of the $N$ particles such that 
$\calp R=(\rr_{\calp 1},\rr_{\calp 2},\ldots,\rr_{\calp N})$, 
with sign ${\rm sgn}(\calp)$. 
Any permutations can be broken into cycles $\calp=\{C_\nu\}$ where $C_\nu$ is the
number of cycles of length $\nu$ in $\calp$. In the sum over the permutations one
should use a $+1$ for the symmetrization necessary for bosons and $-1$ for the 
antisymmetrization necessary for fermions, in ${\rm sgn}(\calp)$. For example 
for fermions, an even (odd) 
permutation has ${\rm sgn}(\calp)=+1 (-1)$ and an even (odd) number of exchanges 
of a pair of particles.

On a surface, $d=2$, for impenetrable identical bodies, one can also have 
anyonic statistics \cite{Lerda}. In this case it is necessary to 
consider, more generally,
\bq \label{eq:dmidim}
&&\rho_\nu(R,R';\beta)=\sum_{\alpha\in B_N}{\rm Re}[\chi(\alpha)]
\tilde{\rho}_\alpha(R,R';\beta),\\ \label{eq:chi}
&&\chi(\{\mbox{paths $R(t)$ with $n$ braids among the pairs of single particle paths 
$\rr(t)$}\})=e^{-i\nu n \pi},
\eq
where $B_N$ is the infinite braid group which admits an infinite number of 
unitary one dimensional representations $\chi$, a phase factor, parametrized 
by an arbitrary number $\nu$ which determines the statistics and 
$\tilde{\rho}_\alpha$ is the distinguishable 
density matrix obtained from paths of kind $\alpha$ only. Clearly for 
$\nu=0$ (mod 2) we recover the Bose-Einstein statistics and for $\nu=1$ 
(mod 2) the Fermi-Dirac statistics. So we will be interested in values of 
$0<\nu<1$.

The braid group is the {\sl fundamental group} of the quotient space 
$(S^{2N}-\Delta)/S_N$ where $S^2$ is the (two) sphere, 
$\Delta=\{R~|~\rr_i=\rr_j~~\mbox{for some}~~i\neq j\}$,
and $S_N$ is the group of permutation of $N$ bodies. We then see how 
paths $R(t)$ with different numbers of crossings between single particle
paths $\rr(t)$ belong to different homotopy classes and one cannot be 
deformed continuously into the other. Therefore in order to take care 
of the density matrix of identical impenetrable bodies it is necessary to 
sum over all the topologically disjoint homotopy classes as is done in Eq. 
(\ref{eq:dmidim}). If $\nu$ is rational, i.e. for {\sl fractional statistics},
the phase factor will be periodic in $n$ and for irrational $\nu$, it will 
not.

For a two dimensional electron gas in a transverse magnetic field the role 
of the Landau level \cite{LandauQMnote} {\sl filling factor} is played by 
the statistics $\nu$ \cite{Ortiz1997a,Ortiz1997b} and certain fractional 
values for $\nu$ serve to explain the {\sl fractional quantum Hall effect}. 
In particular R. Laughlin \cite{Laughlin1983} proposed an ansatz for the 
variational ground state of the {\sl jellium} in a transverse magnetic 
field when the filling factor is $\nu=1/p$ with $p$ an odd integer. In 
his construction of the ground state trial wave function the exact analytic 
solution for the partition function of the one component (non quantum)
plasma at a special value of the coupling constant 
\cite{Jancovici81b,Fantoni19a} played a fundamental role. In this work we
will not worry about introducing a magnetic field but we will study the 
case of the {\sl electron gas} where the electrons, fermions, interact 
with a Coulomb pair potential on the manifold (see Section \ref{sec:ie}).

We will now choose a particularly simple Riemannian manifold: the sphere!

\section{The sphere}
\label{sec:sphere}

A {\sl sphere} of radius $a$ is {\sl the} 
(see 1839 Minding theorem and 1899 Liebmann theorem \cite{Gray}) surface, $d=2$,
\footnote{So it is conformally flat as any Riemannian manifold of dimension 
$d\leq 3$.} of constant positive scalar curvature $2/a^2$ so that $\calr=2N/a^2$.
Her metric is
$ds^2=g_{\alpha\beta}dr^\alpha dr^\beta=a^2(d\theta^2+\cos^2\theta\,d\varphi^2)$,
\footnote{Note that in the kinetic action of the path integral Monte Carlo 
calculation it is crucial to use consistently either
$ds^2(\rr,\rr')=g_{\alpha\beta}(\rr)(\rr-\rr')^\alpha(\rr-\rr')^\beta$ or 
$ds^2(\rr,\rr')=g_{\alpha\beta}(\rr')(\rr-\rr')^\alpha(\rr-\rr')^\beta$.}. 
The polar angle $r^1=\theta\in]-\pi/2,\pi/2]$
and the azimuthal angle $r^2=\varphi\in]-\pi,\pi]$ are the contravariant 
coordinates of the position vector $\rr\in\calc$, a point of the sphere, with 
$\calc=]-\pi/2,\pi/2]\times]-\pi,\pi]$ the single particle positions space.
So that $\theta=0$ is the {\sl equator} and $\theta=\pm\pi/2$ are the {\sl poles}. 
On the sphere $\sqrt{g(\rr)}=a^2|\cos\theta|$. 
And both the curvature term and the van Vleck factor, being constant,
simply drop off from the measure of the various observables of Eq. 
(\ref{eq:measureO}).

The position of a particle on the sphere in the three dimensional Euclidean space 
embedding the sphere is
\bq
\left\{\begin{array}{l}
x=a\cos\theta\cos\varphi\\
y=a\cos\theta\sin\varphi\\
z=a\sin\theta\\
\end{array}\right.
\eq
and the particle path on it is $\qq(t)=(x(t),y(t),z(t))$.

The geodesic distance between particles $\rr_i$ and $\rr_j$ is
\bq \label{eq:gd}
s_{ij}=s(\rr_i,\rr_j)=a\arccos\left[\sin(r_i^1)\sin(r_j^1)+
\cos(r_i^1)\cos(r_j^1)\cos(r_i^2-r_j^2)\right],
\eq
whereas the Euclidean distance is
\bq \label{eq:ed}
d_{ij}=d(\rr_i,\rr_j)=a\sqrt{2(1-\hat{\qq}_i\cdot\hat{\qq}_j)}
=2a\sin[\arccos(\hat{\qq}_i\cdot\hat{\qq}_j)/2].,
\eq
where $\hat{\qq}_i=\qq_i/a$ is the versor that from the center of the
sphere points towards the center of the $i$th particle. 

We use the Metropolis algorithm \cite{Metropolis,Kalos-Whitlock} to evaluate
the average of Eq. (\ref{eq:measureO}). A code listing is offered in the 
supplementary material \cite{suppl} (see also reference [1] therein).

In order to explore ergodically the
positions space $\rr=(\theta,\varphi)\in\calc$ to sample the 
distinguishable density matrix we use the transition 
{\sl displacement} move described in Appendix \ref{app:dmove}.

In order to sample the permutation sum of Eq. (\ref{eq:dmid}) needed for 
identical bodies we use a transition move combination of 2 Brownian 
{\sl bridges} between unlike bodies as described in Appendix \ref{app:bmove}. 
To construct a single Brownian bridge it is essential to map, project, the
sphere on a flat coordinate system. Our choice is presented in Appendix 
\ref{app:bmove}. Perform the Gaussian bridge move in 
the projection flat space and then map it back to the sphere. The Metropolis 
algorithm will then allow to sample the high temperature density matrix 
whose kinetic action is not purely Gaussian on the sphere, due to the 
metric tensor appearing in $d\tilde{s}^2$
\footnote{We tried to avoid passing back and forth through the 
projection mapping but with no results. In fact in order
to be able to construct a Brownian bridge it is necessary to connect 
particles several timeslices apart. And their geodesic distance $s_{ij}$
does not reduce to a simple quadratic form.}. 

In order to sample the sum over the homotopy classes 
of Eq. (\ref{eq:dmidim}) needed for identical (impenetrable) bodies we use a 
combination of bridge and displacement transition moves as described in Appendix 
\ref{app:cmove}. Note that the displacement moves can be freely substituted by 
moves of bridges connecting only like bodies. But we found it convenient to
use both moves for three reasons: (i) a Monte Carlo method usually becomes more
efficient if implemented through a rich menu of different moves; (ii)
from a purely formal point of view, one starts from simple single bead 
moves and only later builds more complex many beads moves; (iii) since the 
single bead displacement move is simple to construct 
it can serve as a test for more elaborated many beads moves.

We will work in the canonical ensemble with fixed number of particles
$N$, surface area $A=4\pi a^2$, surface density $\sigma=N/4\pi a^2$, and 
absolute temperature $T=1/k_B\beta$. In our simulations we will only
consider the $m=1$ case. The many body system
{\sl degeneracy parameter} is $\Theta=T/T_D$ where the
degeneracy temperature $T_D=\sigma\hbar^2/mk_B$
\footnote{This is the temperature at which the size of a single 
particle path, in absence of interaction, i.e. the thermal 
wavelength $(2\lambda\beta)^{1/2}$, equals the separation between 
single particle paths, i.e. roughly $\sigma^{-1/2}$. Below this
temperature it is possible for single particle paths to link up
exchanging their end points.}. For temperatures
higher than $T_D$, $\Theta\gg 1$, quantum effects are not very important
and the distinction between distinguishable particles, bosons, fermions,
or anyons is lost.
We will treat both the non interacting fluid $V=0$ and the Coulomb
fluid
\bq \label{eq:coulomb}
V(R)=\sum_{i<j}\frac{e^2}{d_{ij}},
\eq
where $e$ is the unit of charge and we are assuming that the particles,
moving on the sphere, interact with the three dimensional pair Coulomb 
potential in terms of the Euclidean distance (\ref{eq:ed}) between two
electrons
\footnote{Note that this is not the only possible choice since we could
as well choose particles ``living in'' \cite{Abbott} the surface of the 
sphere as done for example in Ref. \cite{Fantoni03a,Fantoni08c,Fantoni12b}
(for other surfaces)
or particle ``moving on'' the sphere but interacting with the two 
dimensional logarithmic Coulomb potential with the Euclidean distance as 
done for example in Ref. \cite{Caillol81,Fantoni19a}.}.
The {\sl Coulomb coupling constant} is $\Gamma=\beta e^2/a_0r_s$ with
$a_0=\hbar^2/me^2$ the Bohr radius and $r_s=(4\pi\sigma)^{-1/2}/a_0$
the Wigner-Seitz radius. At weak coupling, $\Gamma\ll 1$, the 
plasma becomes weakly correlated.
Choosing length in units of the Wigner-Seitz radius, $a_0r_s=a/\sqrt{N}$, 
and energy in units of Rydberg, $\text{Ry}=\hbar^2/2ma_0^2$, we have
$\lambda=1/r_s^2$, $\Gamma=\beta(2/r_s)$, and $\Theta=(2\pi r_s^2)/\beta$. 
We then see that the kinetic energy scales as $1/r_s^2$ and the potential 
energy as $1/r_s$ so that the electron gas will approach the ideal gas
limit at low $r_s$ or high surface densities $\sigma$.
In this work we will instead choose everywhere natural units, 
with $\hbar=k_B=1$.

Apart from thermodynamic properties like the {\sl kinetic energy} per 
particle $e_K$ and the {\sl potential energy} per particle $e_V$ 
\footnote{The estimators for these observables are carefully described in 
Ref. \cite{Ceperley1995}.} we will also measure structural properties 
like the {\sl radial distribution function}, $g(r)=\langle\calo\rangle$. 
For it we may use the following histogram estimator,
\bq \label{eq:rdf}
O(R;r)=\sum_{i\neq j}
\frac{1_{]r-\Delta/2,r+\Delta/2]}(d_{ij})}{Nn_{id}(r)},
\eq
where $\Delta$ is the histogram bin, $1_{]a,b]}(x)=1$ if $x\in]a,b]$
and 0 otherwise, and 
\bq
n_{id}(r)=N\left[\left(\frac{r+\Delta/2}{2a}\right)^2-
\left(\frac{r-\Delta/2}{2a}\right)^2\right]~,
\eq
is the average number of particles on the spherical crown
$]r-\Delta/2,r+\Delta/2]$ for the ideal gas of density
$\sigma$. We have that $\sigma^2g(r)$ gives the probability, that
sitting on a particle at $\rr$, one finds another particle at
$\rr^\prime$ with $r=d(\rr,\rr')\in[0,2a]$. Note that $\sqrt{2a^2}$
is the Euclidean distance between a pole and a point on the equator. 
Alternatively one could consider a radial distribution
function defined through the geodesic distance $r=s(\rr,\rr')\in[0,\pi a]$
where one chooses $1_{]r-\Delta/2,r+\Delta/2]}(s_{ij})$ in
Eq. (\ref{eq:rdf}) with an appropriate normalization
\footnote{In Ref. \cite{Fantoni18c} we used the Euclidean distance.}.

For $e_K$ and $e_V$ we will use the {\sl direct} estimator described in 
Ref. \cite{Ceperley1995} applied to our action of Eq. (\ref{eq:action}). 
So that the kinetic energy per particle 
$e_K=\langle K(R_k,R_{k+1};\tau)\rangle/N$ with 
$K(R,R';\tau)=N/\tau-\tilde{g}_{\mu\nu}(R)(R-R')^\mu(R-R')^\nu/4\lambda\tau^2$ 
and the physical potential 
energy per particle $e_V=\langle V(R_k)\rangle/N$. 
During the simulation we will also measure the effective geometrical 
potential energy per particle $e_W=\langle W(R_k)\rangle/N$.
For bosons we will also measure the {\sl superfluid fraction} from the 
{\sl area estimator} of Eqs. (\ref{eq:sfarea})-(\ref{eq:sfarea1}).

In Table \ref{tab:tq} we list some case studies treated in our computer
experiments. In all cases we use $M=50$ timeslices and a surface density 
$\sigma=1/10\pi$. 

\begin{table*}[htbp]
\protect\caption{Cases treated in our simulations:
  $a$ the sphere radius,  
  $N$ the number of particles, $\beta$ the inverse temperature,  
  $e_K=\langle K\rangle/N$ the kinetic
  energy per particle from the thermodynamic estimator as explained
  in Ref. \cite{Ceperley1995}, $e_V=\langle V\rangle/N$ the physical potential 
  energy per particle, and $e_W=\langle W\rangle/N$ the geometrical 
  potential energy per particle. 
  The other quantities were introduced in the 
  main text. In the statistics column ``d'' stands for distinguishable, 
  ``b'' for bosons, ``f'' for fermions, and ``a'' for anyons. The 
  statistics $\nu\in[0,1]$ is also reported. We chose natural units 
  with $\hbar=k_B=1$. The simulation lasted more than $10^8$ Monte Carlo steps
  where one step is made of a displace move of all the beads of a single
  particle path as described in Appendix \ref{app:dmove} and a bridge 
  move as described in Appendix \ref{app:bmove}. Case F was the most
  computationally expensive and took roughly 4 weeks of computer time.}  
\vspace{1cm}
\label{tab:tq}
\begin{ruledtabular}
\begin{tabular}{|c|lllllllllll|}
\hline
case & statistics & $M$ & $N$ & $a$ & $a_0r_s$ & $\beta$ & $\Gamma$ & $\Theta$ & $Ne_K$ & $Ne_V$ & $-Ne_W$\\ 
\hline
A & d           & 50 & 10 & 5 & $\sqrt{5/2}$ & 10    & 0 & $\pi$     & 1.094(3) & 0 & 148.39(5)\\
B & b $\nu=0$   & 50 & 10 & 5 & $\sqrt{5/2}$ & 100/3 & 0 & $3\pi/10$ & 0.222(2) & 0 & 44.46(1)\\
C & f $\nu=1$   & 50 & 10 & 5 & $\sqrt{5/2}$ & 100/3 & 0 & $3\pi/10$ & 0.901(4) & 0 & 44.47(1)\\
D & a $\nu=1/2$ & 50 & 10 & 5 & $\sqrt{5/2}$ & 100/3 & 0 & $3\pi/10$ & 0.848(5) & 0 & 44.43(1)\\
E & a $\nu=1/3$ & 50 & 10 & 5 & $\sqrt{5/2}$ & 100/3 & 0 & $3\pi/10$ & 0.945(5) & 0 & 44.64(1)\\
F & f $\nu=1$   & 50 & 15 & $5\sqrt{3/2}$ & $\sqrt{5/2}$ & 100/3 & $20\sqrt{10}~e^2/3$ & $3\pi/10$ & 1.655(4) & 13.984(1) & 75.232(8)\\
G & f $\nu=1$   & 50 & 10 & 5             & $\sqrt{5/2}$ & 100/3 & $20\sqrt{10}~e^2/3$ & $3\pi/10$ & 1.133(4) & 7.070(1) & 44.202(8)\\
H & f $\nu=1$   & 50 & 5  & $5/\sqrt{2}$  & $\sqrt{5/2}$ & 100/3 & $20\sqrt{10}~e^2/3$ & $3\pi/10$ & 0.543(3) & 2.0572(7) & 16.952(8)\\
$\ldots$&&&&&&&&&& &\\
\hline
\end{tabular}
\end{ruledtabular}
\end{table*}
%

\section{Non interacting bodies}
\label{sec:ni}

Here we will study non interacting bodies with 
$\Gamma=0$ or more generally $V=0$. We are not aware of any analytic exact
solution even for this simple case on the sphere. Whereas the flat case
can be studied in closed form, but only in the grand canonical ensemble
\cite{FeynmanFIP,Fantoni24d}. \red{The non interacting quantum many body 
system on the sphere is equivalent to a non interacting quantum many body 
system on a plane under the influence of an effective external potential 
due to the curvature of the sphere \cite{Bastianelli2017}.}

\subsection{Distinguishable bodies}
\label{sec:db}

Here we use the Metropolis algorithm with the transition displacement moves 
of Eqs. (\ref{eq:tdisp}) and (\ref{eq:pdisp}) or equivalently bridge moves 
of Eqs. (\ref{eq:bmove1}) and (\ref{eq:bmove2}) but between like particles.

\begin{figure}[htbp]
\begin{center}
\includegraphics[width=8cm]{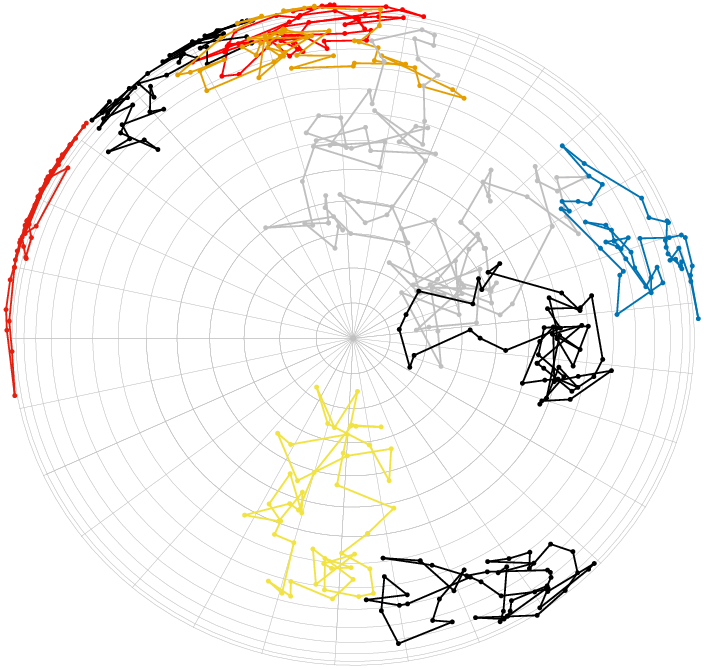}
\includegraphics[width=8cm]{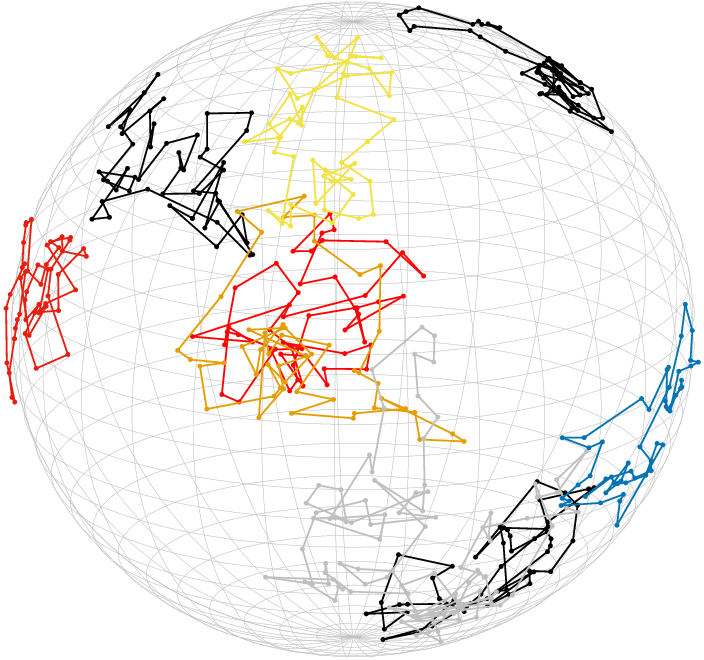}
\end{center}  
\caption{Snapshot of the macroscopic path during the
  simulation for $N=10$ non interacting distinguishable particles 
  with $M=50, a=5, \beta=10$. Case A in Table \ref{tab:tq}.
  The simulation started with all bodies distributed uniformly 
  randomly on the equator.
  The different paths have different colors. In the 
  left panel the top view and in the right panel the 
  front view. In the simulation we measured 
  $e_K=\langle\calk\rangle=1.089(4)$. Reducing $\beta$ each 
  path shrinks and tends to form a ring enclosing less amount 
  of area.}
\label{fig:pathd}
\end{figure}

In Fig. \ref{fig:pathd} we show a snapshot during the simulation for 
distinguishable non interacting particles with $M=50, N=10, a=5, \beta=10$. 

Many snapshots of the paths configurations during the simulation showed that 
the simulation ``speed'' of the beads of the paths near the poles gets small
\footnote{In fact we have two different contributions responsible for this
behavior: The $g$ factor in the integration measure and the metric $g_{\mu\nu}$
in the kinetic action. These cannot be removed with a change of 
coordinates since it would require a {\sl non coordinate basis} 
\cite{Gravitation}.}. This is a consequence of the hairy ball theorem 
mentioned in the introduction. In fact each link of the single 
particle paths $\rr(t)$ belongs to the tangent bundle of the
sphere.

\begin{figure}[htbp]
\begin{center}
\includegraphics[width=10cm]{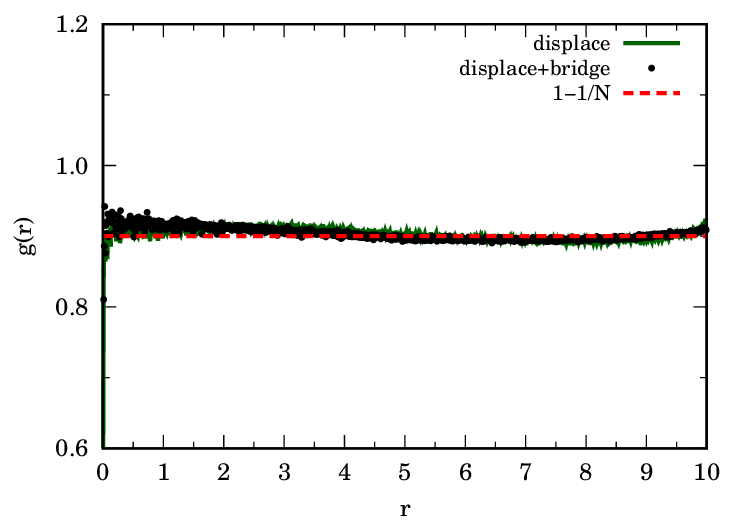}
\end{center}  
\caption{The radial distribution function for the non interacting 
distinguishable particles gas with $N=10$ on a sphere of radius $a=5$ 
at an inverse temperature $\beta=10$. We use $M=50$ (case A in Table 
\ref{tab:tq}) with only the displace move
of appendix \ref{app:dmove} and with both the displace and the bridge 
move as described in Appendix \ref{app:bmove}. 
The dashed line is for $g(r)=1-1/N=0.9$.}    
\label{fig:dgr}
\end{figure}

In Fig. \ref{fig:dgr} we show the radial distribution function, calculated 
by averaging the estimator of Eq. (\ref{eq:rdf}), 
for a gas of $N=10$ distinguishable non interacting particles on a sphere 
of radius $a=5$ at an inverse temperature $\beta=10$. We used $M=50$ with
only the displace move of appendix \ref{app:dmove} (this was used for Fig. 
\ref{fig:pathd}) and with both the displacement move and the bridge move
of Appendix \ref{app:bmove}. From the figure we see how both the simulation 
with only the displacement move and the one with both the displacement and 
the bridge move give the expected result $g(r)=1-1/N$, where the $1/N$ term 
takes care of the finite size of the system, for any $M$. This is a strong 
test on the correctness of our bridge move. We are then ready to use it 
for the permutations sampling necessary for identical particles. We will
do this in the next section.

%

\subsection{Identical bodies}
\label{sec:ib}

Here we use the Metropolis algorithm with the transition displacement moves 
of Eqs. (\ref{eq:tdisp}) and 
(\ref{eq:pdisp}) and bridge moves of Eqs. (\ref{eq:bmove1}) and 
(\ref{eq:bmove2}) between unlike particles as described in Appendix 
\ref{app:bmove} to produce the necessary particles exchanges.

\subsubsection{Bosons}

Given the {\sl superfluid} density $\sigma_s$ and the {\sl normal fluid} density
$\sigma_n=1-\sigma_s$, the area estimator for the superfluid fraction is given 
by \cite{Pollock1987,Ceperley1995}
\bq \label{eq:sfarea}
f_s=\frac{\sigma_s}{\sigma}=1-\frac{\sigma_n}{\sigma}=
\frac{2m\langle \cala^2\rangle}{\beta\lambda I_c},
\eq
where, if $\epsilon$ is the Levi-Civita antisymmetric symbol,
\bq \label{eq:sfarea1}
\cala=\frac{1}{2}\sum_{i,k}\epsilon_{\alpha\beta}
(r_{i,k+1}-r_{k})^\alpha 
(r_{i,k+2}-r_{k+1})^\beta\sqrt{g(\rr_{i,k+1})},
\eq
is the area enclosed by all the single particle paths and 
$I_c$ is the classical moment of inertia of the spherical shell
that we will take as a fit parameter so that 
$\lim_{\beta\to\infty}f_s=1$. \red{Therefore our $f_s$ is not an 
absolute superfluid fraction but a relative one. Moreover we adapted 
to a curved surface the area estimator (\ref{eq:sfarea}) that 
Ceperley and Pollock (see section III.E of Ref. \cite{Ceperley1995} and 
Eq. (3.26) therein) derived for the Euclidean flat case. We did this by 
using for the area enclosed by all the single particle paths the
expression (\ref{eq:sfarea1}) valid for paths on the curved 
Riemannian surface. Our estimator (\ref{eq:sfarea}) is
an exact fluctuation-dissipation formula. Superfluidity is a microscopic 
property that can be defined in a finite system. 
It is not necessary to take the thermodynamic limit 
or to have a phase transition to see its effect. In the high temperature
limit $f_s\to 0$ since the paths become short and straight and in the 
low temperature limit $f_s\to 1$ since the paths are long and the 
thermal average in $\sigma_n$ sums up to zero.}

In Ref. \cite{Nelson1977} Nelson and Kosterlitz use renormalization method
of Ref. \cite{Jose1977} to study the behavior of the superfluid density defined
in Ref. \cite{Feynmann1955} at the superfluid phase transition. They found 
that the superfluid density $\sigma_s$ undergoes a {\sl universal} 
jump equal to,  
\bq \label{eq:rhosu}
\Delta\sigma_s=\frac{m^2}{\beta}\frac{2}{\pi},
\eq
at the critical temperature, $T_c$, for the superfluid phase transition.
This was also observed experimentally \cite{Bishop1978} for $^4$He films. 
In the present language this says that the average squared area has a
jump of $(I_c/\sigma)/2\pi$ at the transition. Naturally the phase transition
can only occur in the thermodynamic limit \cite{Tononi2022a} which in our 
case would correspond to the case of a degenerate sphere of an infinite 
radius, i.e. flat.

\begin{figure}[htbp]
\begin{center}
\includegraphics[width=10cm]{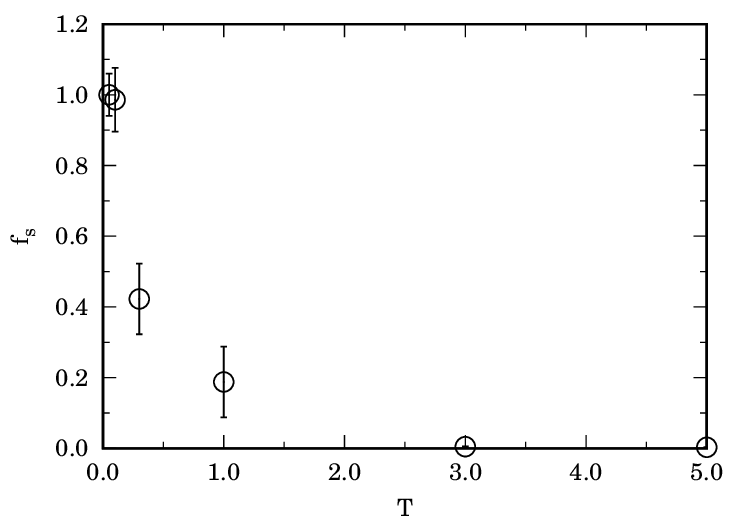}
\end{center}  
\caption{The supefluid fraction (\ref{eq:sfarea}) for the condensate of
non interacting bosons with $N=10,a=5,M=50$. In this case $\tau\leq 0.4$ and
$T_c=T_D\approx 0.0318$. The universal jump in $f_s$ which would be expected 
at the superfluid phase transition is $(4m^2T_c/\sigma)/2\pi\approx 0.636$.}    
\label{fig:sf}
\end{figure}

In Fig. \ref{fig:sf} we show the superfluid fraction of Eq. 
(\ref{eq:sfarea}) for the condensate of non interacting bosons with 
$N=10, a=5$, $M=50$. From the figure we see how the transition from 
$f_s=0$ at high temperature to $f_s=1$ at low temperature occurs in a
temperature interval $\Delta T\approx 3$ and $f_s\approx 1$ is reached 
at the critical condensation temperature $T\approx T_D=0.0318$ as expected. 

\begin{figure}[htbp]
\begin{center}
\includegraphics[width=8cm]{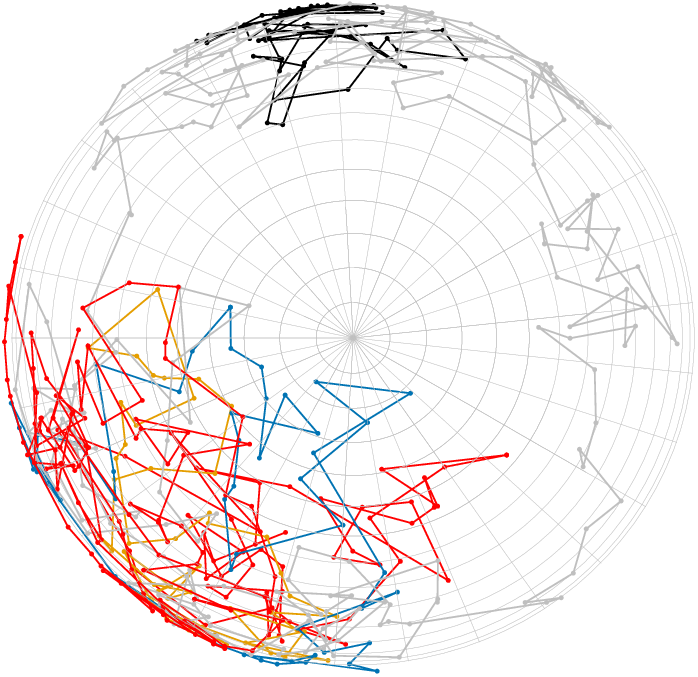}
\includegraphics[width=8cm]{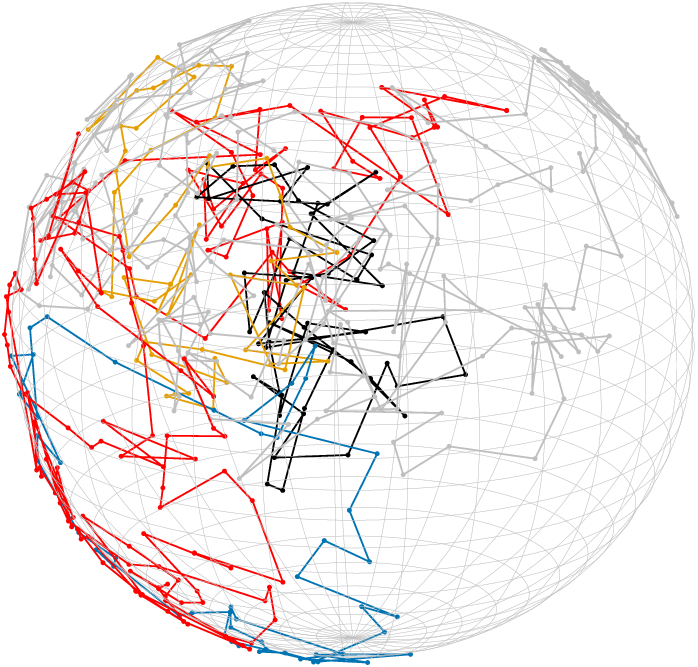}
\end{center}  
\caption{Snapshot of the macroscopic path during the
  simulation for $N=10$ non interacting bosons
  with $M=50, a=5, \beta=100/3$. Case B in Table \ref{tab:tq}.
  The simulation started with all bodies distributed uniformly 
  randomly on the equator.
  Paths corresponding to different permutations cycles have 
  different colors. In the 
  left panel the top view and in the right panel the 
  front view. In the simulation we measured 
  $e_K=\langle\calk\rangle=0.224(2)$. Reducing $\beta$ each 
  path shrinks and tends to form a ring enclosing less amount 
  of area.}
\label{fig:pathb}
\end{figure}

In Fig. \ref{fig:pathb} we show a snapshot during the simulation for 
bosons non interacting particles with 
$M=50, N=10, a=5, \beta=100/3$.
We see how the system forms 5 permutation cycles corresponding to 
different colors.

In Fig. \ref{fig:bfgr} we show the radial distribution function, calculated 
by averaging the estimator of Eq. (\ref{eq:rdf}), 
for the bosons non interacting gas with $N=10$ on a sphere of radius 
$a=5$ at an inverse temperature $\beta=100/3$ with $M=50$. This is the 
same configuration used for Fig. \ref{fig:pathd} but at a lower temperature. 
The bump at $r=0$ is a 
consequence of condensation predicted by the Bose-Einstein statistics which
requires a symmetric density matrix respect to permutaion of any 
two particles. Bosons like themselves \cite{Fantoni21i} but on a 
sphere if they like themselves near contact they must form a hole on the
opposite point.

\subsubsection{Fermions}

Fermions properties cannot be calculated exactly with path integral
Monte Carlo because of the {\sl fermions sign problem}
\cite{Ceperley1991,Ceperley1996}. We then have to resort to an
approximate calculation. The one we chose was the Restricted Path
Integral (RPIMC) approximation \cite{Ceperley1991,Ceperley1996,Fantoni18c,Fantoni21i} 
with a ``free fermions restriction''. The trial density matrix used in the
{\sl restriction} is chosen as the one reducing to the ideal density matrix
in the limit of $t\ll 1$ and is given by the following explicit analytic 
expression,
\bq \label{ifdm}
\rho_0(R',R;t)\propto{\rm det} \left|\left|e^{
-\frac{s^2(\rr_i',\rr_j)}
{4\lambda t}}\right|\right|.
\eq
Note that here we need the geodesic distance between two particles 
on the sphere, $s^2(\rr',\rr)$, that in the $t\ll 1$ limit reduces 
to $ds^2(\rr',\rr)=g_{\alpha\beta}(\rr)(\rr-\rr')^\alpha
(\rr-\rr')^\beta$.

The {\sl restricted path integral identity} that we will use states
\cite{Ceperley1991,Ceperley1996} 
\bq \label{rpii}
\rho_-(R',R;\beta)&\propto&\int \sqrt{\tilde{g}''}
dR''\,\rho_-(R'',R;0)\tilde{\rho}_{\gamma_0(R'')}(R',R'';\beta),
\eq
where $\tilde{\rho}_{\gamma_0(R'')}(R',R'';\beta)$ is the 
distinguishable density matrix and the subscript means that we 
restrict the path integration in (\ref{eq:pi}) to 
paths starting at $R''$, ending at $R'$, and avoiding the nodes of
$\rho_0$, i.e. to the {\sl reach} of $R''$, $\gamma_0(R'')$. 
$R''$ will be called the {\sl reference point} determining the reach. 
The nodes are on the reach boundary $\partial\gamma_0$. The weight of
the walk is $\rho_-(R'',R;0)={\rm det}||\delta(\rr_i''-\rr_j)||$. 
Note that, in imposing the restriction it is convenient to 
imagine an infinitely positive external potential which will
prevent a transition move $R\to R'$ such that 
$\rho_0(R',R;\tau)<0$
\footnote{So that a move that changes the sign of $\rho_0$ are 
rejected in the Metropolis scheme.}. It is clear that the  
contribution of all the paths for a single element of the density
matrix will be of the same sign, thus solving the sign problem;
positive if $\rho_-(R'',R;0)>0$, negative otherwise. On the
diagonal any density matrix is positive and on the path restriction
$\rho_-(R',R;\beta)>0$, then only even permutations, those with
${\rm sgn}(\calp)=+1$, are allowed, since 
$\rho_-(\calp R,R;\beta)={\rm sgn}(\calp)\rho_-(R,R;\beta)$. It
is then possible to use a bosons calculation to get the fermions
case. Clearly the restricted path integral identity with the free
fermions restriction becomes exact if we simulate free fermions, but
otherwise is just an approximation. 

The restriction implementation is rather simple: we
just reject the move whenever the proposed path is such
that the ideal fermion density matrix (\ref{ifdm}) calculated between
the reference point and any of the time slices subject to
newly generated particles positions has a negative sign.
So in correspondence of each displace move of Appendix \ref{app:dmove} 
or bridge move of Appendix \ref{app:bmove} it is necessary to 
calculate $M$ determinants of order $N$. For this reason the method
becomes unfeasible to treat systems with very many particles.
To increase the acceptances in the restrictions, we found it 
convenient to choose the reference point time 
slice randomly, i.e. we choose an integer random number between $1$
and $M$, say $m$, and the reference point is chosen to be $R=R_m$, before 
each move. This is allowed because we are free to perform a translation 
in the $\beta$-periodic imaginary thermal time.

\begin{figure}[htbp]
\begin{center}
\includegraphics[width=10cm]{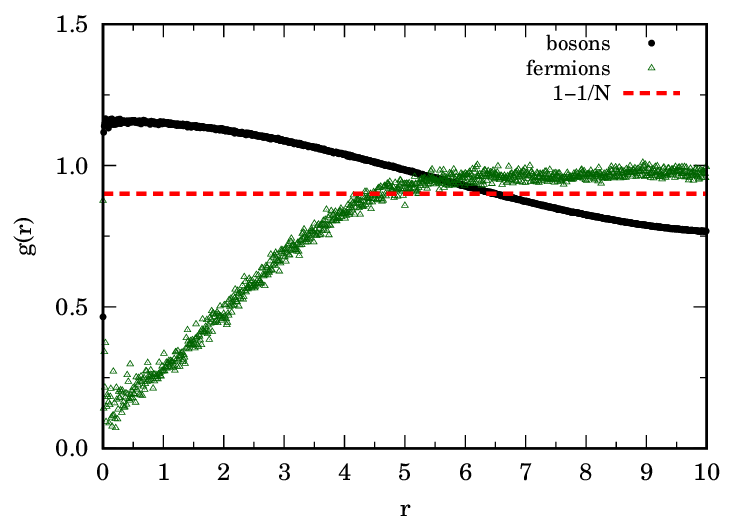}
\end{center}  
\caption{The radial distribution function for 
the non interacting bosons and fermions gas with $N=10$ on a sphere of radius 
$a=5$ at an inverse temperature $\beta=100/3$ with $M=50$. Case B and C in Table 
\ref{tab:tq} respectively. The dashed line is for $g(r)=1-1/N=0.9$. The bump at 
$r=0$ for bosons is a manifestation of their tendency to like themselves. The 
exchange hole at $r=0$ for fermions is a manifestation of their tendency to 
dislike themselves due to the Pauli exclusion principle and it requires 
$g(0)=0$. \red{The fact that $g(r)$ seems to approach not exactly 0 as $r\to 0$
is due to the path integral discretization error and we explicitly verified 
that choosing smaller timesteps $\tau$, $g(0)$ comes even closer to 0.}}    
\label{fig:bfgr}
\end{figure}

In Fig. \ref{fig:bfgr} we show the radial distribution function, calculated 
by averaging the estimator of Eq. (\ref{eq:rdf}), for the non interacting, 
fully polarized fermions gas with $N=10$ on a sphere of 
radius 
$a=5$ at an inverse temperature $\beta=100/3$ with $M=50$. This is case
C of Table \ref{tab:tq}. The well at $r=0$ is a consequence of Pauli
exclusion principle predicted by the Fermi-Dirac statistics which
requires an antisymmetric density matrix respect to permutaion of any 
two particles. This well is usually called {\sl exchange hole}. Fermions 
dislike themselves \cite{Fantoni21i} but on a 
sphere if they dislike themselves at one point they must form a bump on the
opposite point. A simple sum rule in this case requires $g(0)=0$ since the
the density matrix for coincident particles is singular and its 
determinant must be zero. 

%

\subsubsection{Anyons}
\label{sec:ab}

Here we use the Metropolis algorithm with the transition displacement 
moves of Eqs. (\ref{eq:tdisp}) and (\ref{eq:pdisp}) and the swap moves of 
Eqs. (\ref{eq:bmove1}) and (\ref{eq:bmove2}) between two particles, as 
described in Appendix \ref{app:cmove}, counting at each move the number 
of single particles crossings $n$ in the newly generated path $R$.

For example, from Eqs. (\ref{eq:dmidim})-(\ref{eq:chi}) follows that, for 
$\nu=1/2$,
\bq \label{eq:nu1/2}
{\rm Re}[\chi(\alpha)]=\left\{\begin{array}{ll}
(-1)^k & n=2k\\
0      & n=2k+1
\end{array}\right.~~~k=0,1,2,3,\ldots
\eq
We then see that in order to calculate $\rho_{1/2}$ necessary to simulate 
anyonic statistics for $\nu=1/2$, one can simply use the RPIMC described 
above, in order to get rid of the sign problem, i.e. a bosonic PIMC with 
a restriction based on the nodes of the reference trial density matrix of 
Eq. (\ref{ifdm}), as for fermions. But now, after the 
Metropolis acceptance, we simply have to additionally 
throw away those moves that generate an odd number $n$ of single particles 
crossings, i.e. when $n=1$ (mod 2). In order to count the number $n$ of
braids we can proceed as described in Appendix \ref{app:cmove}. 
Once again, as long as we have no interaction between 
the anyonic particles this scheme is expected to give rise to an exact 
computation. 

On the other hand we suspect that fractional statistics $\nu=q/p$ with 
$0<q<p$ and $p>2$ will give rise to fluids with different structure. For 
example, for $\nu=1/3$ we find
\bq \label{eq:nu1/3}
{\rm Re}[\chi(\alpha)]=\left\{\begin{array}{ll}
(-1)^k                 & n=3k\\
(-1)^k\cos(\pi/3)      & n=3k+1\\
(-1)^k\cos(2\pi/3)     & n=3k+2
\end{array}\right.~~~k=0,1,2,3,\ldots
\eq
and the $\rho_{2/3}$ should differ from the $\rho_{1/3}$. To simulate
the $\nu=1/3$ case we have to use again the RPIMC, in order to get rid of the 
sign problem, but now, differently from the $\nu=1/2$ case, we 
have to weight each $\tilde{\rho}_\alpha$ with the correct factor, according to
(\ref{eq:nu1/3}): $1$ for $n=3k$ and $1/2$ otherwise. So that, after the 
Metropolis acceptance, if $n=0$ (mod 3) we do nothing and if $n\neq 0$ 
(mod 3) we additionally throw away the moves with probability $1/2$.

\begin{figure}[htbp]
\begin{center}
\includegraphics[width=10cm]{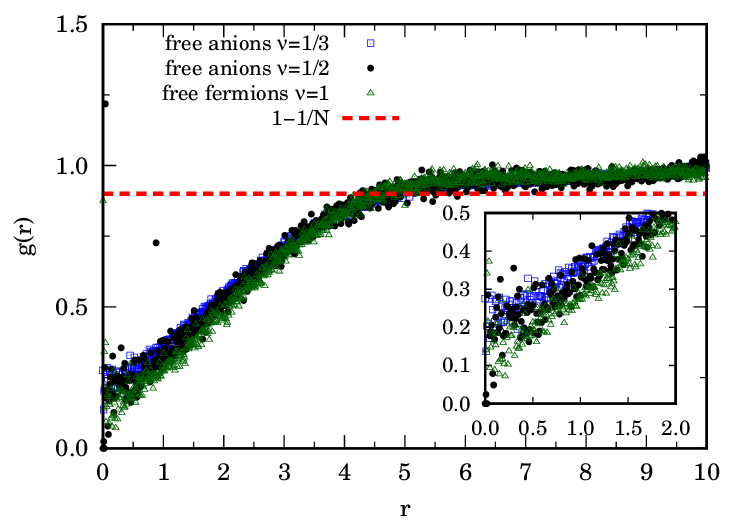}
\end{center}  
\caption{The radial distribution function for 
the non interacting $\nu=1/2$ and $\nu=1/3$ anyons gas with $N=10$ on a 
sphere of radius $a=5$ at an inverse temperature $\beta=100/3$ with $M=50$. 
Case D and E in Table \ref{tab:tq} respectively. The dashed line is for 
$g(r)=1-1/N=0.9$. From the inset we see how the structure of the free 
$\nu=1/2$ anyons 
develops a slightly smaller exchange hole than fermions and the exchange 
hole of the free $\nu=1/3$ anyons is slightly smaller than the one of the
free $\nu=1/2$ anyons.}    
\label{fig:agr}
\end{figure}

In Fig. \ref{fig:agr} we show the radial distribution function, calculated 
by averaging the estimator of Eq. (\ref{eq:rdf}), for the non interacting 
$\nu=1/2$ and $\nu=1/3$ anyons gas with $N=10$ on a sphere of radius 
$a=5$ at an inverse temperature $\beta=100/3$ with $M=50$. These are cases
D and E of Table \ref{tab:tq}. We compare them with the radial distribution 
for non interacting fermions already shown in Fig. \ref{fig:bfgr}. From the 
comparison we can say that we observe a slight diminution of the exchange 
hole as we move from (fermions) $\nu=1$ to (anyons) $\nu=1/2$ to (anyons) 
$\nu=1/3$. Moreover we find
$g_{1/3}(0)>g_{1/2}(0)>g_-(0)=0$, where we indicate with $g_\nu$ the 
radial distribution function for statistics $\nu$. The kinetic energy of the
(anyons) $\nu=1/2$ case is smaller than the one for $\nu=1$ (fermions) and 
the one of the $\nu=1/3$ case is bigger than the one for $\nu=1$.

It would also be interesting to explore other rational cases $\nu=q/p$ with 
$0<\nu<1$ and see whether their structure differs in any way from the 
irrational cases.

\section{Electron gas}
\label{sec:ie}

In this section we will study an electron gas, i.e. fermions interacting 
through the Coulomb potential of Eq. (\ref{eq:coulomb}) with $e=1$.
We will only consider the fully polarized case where the density
matrix is antisymmetric respect to permutation of any two particles.
In this case we are unable to solve exactly the problem not even numerically 
with PIMC due to the fermions sign problem described above.
We will then use the RPIMC with the ideal free fermions restriction based on
the nodes of the reference density matrix of Eq. (\ref{ifdm}). This
strategy, which is exact for non interacting fermions, is here expected to
become a better approximation at high density and high temperature, i.e. 
when correlation effects are weak. 
\footnote{\red{Note that in our implementation 
of the RPIMC we just added the approximate free particle restriction 
to the exact PIMC for bosons. Doing like so we are neglecting almost 
all even permutations since our PIMC samples the permutations through 
the combination of two bridge and a swap moves as described in Appendix 
\ref{app:bmove} but the restriction rejects almost all exchanges of two
particles that would change the sign of the free particle density 
matrix (\ref{ifdm}). In order to fully sample the even
permutations in the RPIMC it is then necessary to add to our moves menu' 
a third move which realizes a cyclic permutations of 3 particles
$123\to 312$. This will have no effect in the non interacting case but
is expected to become relevant in the interacting case. Here we simply 
neglected this third move even in the interacting case.}}

We simulated cases F, G, H of Table \ref{tab:tq} where we keep constant the
thermodynamic properties of the electron gas of temperature, $T=3/100$,
and surface density, $\sigma=1/10\pi$, and increase gradually the radius 
of the sphere $a=5/\sqrt{2}, 5, 5\sqrt{3/2}$. From a comparison of the
results for the thermodynamic properties of the many body system we see
that the kinetic energy per particle remains approximately constant 
whereas the potential energy per particle diminishes as the curvature
increases.

\begin{figure}[htbp]
\begin{center}
\includegraphics[width=10cm]{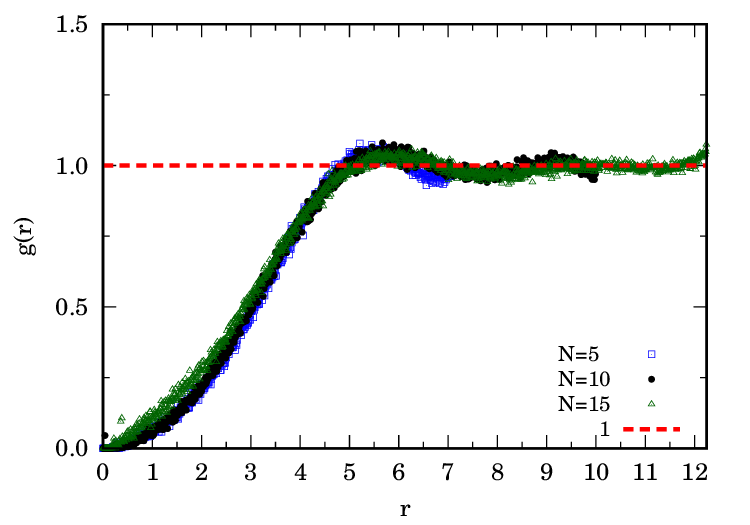}
\end{center}  
\caption{The radial distribution function for the fully polarized 
electron gas interacting through the Coulomb potential of Eq. 
(\ref{eq:coulomb}) with $e=1$. We show cases F, G, and H of Table \ref{tab:tq} 
at constant $\sigma=1/10\pi$ and $N=5,10,15$ electrons respectively, at an 
inverse temperature $\beta=100/3$ with $M=50$. As the curvature 
increases at constant density, near contact the exchange-correlation 
hole tends to increase its extend and the long range oscillations around 
$g(r)=1$ tend to to curl either upward or downward in a neighborhood of 
the ending point at $r=2a$.}    
\label{fig:egr}
\end{figure}

In Fig. \ref{fig:egr} we show the radial distribution function for the
electron gas of cases F, G, and H in Table \ref{tab:tq} where we keep
the surface density constant changing the number of particles $N=5,10,15$
respectively. From the figure we see how
the gas develops a {\sl correlation hole} at $r=0$ in addition to the 
exchange hole already shown in Fig. \ref{fig:bfgr}. The radial distribution 
function raises from $g(0)=0$ at contact developing oscillations beyond a
certain distance $r\gtrsim r_N$ around $g=1$.
From the comparison between cases F, G, and H we see how the 
{\sl exchange-correlation hole} tends to increase as the curvature 
increases at constant surface density and at the same time 
$r_{5}\approx r_{10}\approx r_{15}\approx 4.5$ and the oscillations 
tend to curl either upward or downward in a neighborhood of $r=2a$.

%

\section{Conclusions}
\label{sec:conclusions}

In this work we studied the effect of a constant positive curvature on a two
dimensional quantum many body fluid. This is the simplest case that can 
be thought of in an exploration of the influence of curvature on the surface 
where the bodies move. At sufficiently low temperature quantum effects become 
important and the fluid will behave differently depending on the statistics
ruling the bodies. The statistics depends on the transformation property of
the many body wavefunction under exchange of two bodies. If the bodies
are distinguishable the wavefunction and its transformed will be different, 
if they are identical the transformed wavefunction can only be plus or
minus the original wavefunction. The plus will be for bodies obeying to 
Bose-Einstein statistics and the minus for bodies obeying to the Fermi-Dirac
statistics. If the identical bodies are also impenetrable the transformed 
wavefunction can more generally be the original wavefunction times a phase
factor $e^{i\nu\pi}$, since, on a surface, exchanging the two particles by a 
clockwise rotation of $\pi$ around their center of mass is a topologically and 
physically distinct operation than rotating counterclockwise. In this case 
the bodies are said to obey to the anyonic statistics, where the statistics is 
determined by the phase $\nu$. For $\nu=0$ (mod 2) one recovers the bosonic 
statistics and for $\nu=1$ (mod 2) the fermionic statistics. 
While for bosons and fermions just the permutation of the particles in their 
initial and final configuration in imaginary time matters, for anyons it is also 
necessary to specify how the different trajectories wind or {\sl braid} around 
each other during their imaginary time evolution. 
In other words the imaginary time evolution of the particles
matters and cannot be neglected. The representation of the permutation group 
must be replaced by the one of the braid group.

We made some computer experiments for each one of these cases using a PIMC
algorithm different from the one previously used in Ref. 
\cite{Fantoni18c,Fantoni23a}. For the sign problem hidden in the 
Fermi-Dirac and the anyonic statistics we then used the RPIMC method.
We studied both the case of the non interacting bodies and for the fermions 
case the electron gas with a three dimensional pair Coulomb interaction 
depending on the Euclidean distance between two electrons. In each case
we measured the kinetic energy, the potential energy, and the radial 
distribution function. The pair correlation function of non interacting 
fermions displays, at contact, the exchange hole and the one of the 
electron gas displays the exchange-correlation hole. While fermions dislike 
themselves due to the Pauli exclusion principle, bosons like themselves and 
form a condensate. This is reflected in a bump in the pair correlation function 
at contact. For bosons we also measured the superfluid fraction using an area
estimator devised by Pollock and Ceperley. Even if on a finite sphere we
clearly cannot reach the thermodynamic limit, the universal jump
predicted by Nelson and Kosterlitz at the suerfluid phase transition 
still gives some indications of the behavior of the superfluid fraction at 
the critical temperature.
 
The effect of curvature is then made manifest by comparing with similar
studies in a flat Euclidean space in 3 dimensions 
\cite{Fantoni21b,Fantoni21i} (compare for example our Fig. \ref{fig:bfgr} 
with Fig. 3 in Ref. \cite{Fantoni21i}) 
or in 2 dimensions \cite{Tanatar1989,Magro1994,Gordillo1998}.
Here we compared the electron gas at the same thermodynamic conditions
of temperature and surface density on spheres of different radii. 
From the results of our computer experiments for cases F, G, and H of 
Table \ref{tab:tq} we see that changing the curvature at constant 
density affects both the extent of the exchange correlation hole at contact
and the oscillations at large distance. As the curvature increases the 
exchange-correlation hole tends to increase and the oscillations tend
to curl either upward or downward at the pole opposite to contact. 
The kinetic energy per particle remains 
approximately constant upon changing the curvature whereas the physical
and geometrical potential energy per particle diminish as the curvature 
increases.

During our
simulations we made various snapshots of the many body path configuration 
and we noticed that the simulation ``speed'' of the beads of the single particle 
paths in proximity of the poles diminishes. We explained this occurrence as a 
consequence of the metric tensor properties which affect both the kinetic 
action and the path integral measure. A consequence of the ``hairy ball
theorem'' which cannot be removed with a change of coordinates.

For the case of anyons we only considered the fractional statistics $\nu=1/2$
and $\nu=1/3$ in the non interacting case. We found that the structure of 
the ideal anyons present an exchange hole that gets smaller moving from 
fermions with $\nu=1$ to anyons with $\nu=1/2$ and from anyons with $\nu=1/2$ 
to anyons with $\nu=1/3$. The kinetic energy of the
$\nu=1/2$ is smaller than the one for $\nu=1$ and the one of 
the $\nu=1/3$ is bigger than the one for $\nu=1$.

It would be interesting to understand whether in the transition from a 
rational to an irrational statistics, there is any macroscopic 
observable change in the structure of the anyons fluid or in his thermodynamic
properties. 

In future computer experiments we plan to measure the effect of curvature on 
the pressure of the fluid as was done analytically in the non quantum case in 
Ref. \cite{Fantoni03a}.

\appendix
\section{The transition displacement move}
\label{app:dmove}

In order to explore the $\theta$ and $\varphi$ positions space 
$\calc=]-\pi/2,\pi/2]\times]-\pi,\pi]$ on the sphere
it is convenient to propose the following displacement transition move for 
each particle in a randomly chosen bead of the many body path
\bq \label{eq:tdisp}
\theta_{\rm new}&=&\theta_{\rm old}+\Delta_\theta(\eta-1/2),\\ \label{eq:pdisp} 
\varphi_{\rm new}&=&\varphi_{\rm old}+\Delta_\varphi(\eta-1/2),
\eq
where $\eta\in [0,1]$ is a uniform pseudo random number and $\Delta_\theta$ 
and $\Delta_\varphi$ are two positive quantities measuring
the $\theta$-displacement and the $\varphi$-displacement respectively.

This transition move can bring $\rr_{\rm new}$ out of $\calc$ so it is also
necessary to bring it back into $\calc$ enforcing periodic boundary conditions
$\varphi=\varphi+2\pi$ and $\theta=\theta+\pi$ with the following subsequent 
chain of transformations
\bq \label{eq:ctrans}
\left\{\begin{array}{l}
\theta_{\rm new}\to\theta_{\rm new}
-\pi~\UseVerb{nint}(\theta_{\rm new}/\pi),\\
\varphi_{\rm new}\to\varphi_{\rm new}
-2\pi~\UseVerb{nint}(\varphi_{\rm new}/2\pi),
\end{array}\right.
\eq
where \verb1NINT1 is the nearest integer function. 
One can easily convince himself that this chain does not alter the 
uniformity of the probability distribution of $\rr_{\rm new}$ in $\calc$.

Note that the metric enters the free particle variance since it is not 
possible by a change of coordinates to remove it both from the 
kinetic-action and from the integration measure $\sqrt{g(\rr)}\,d\rr$,
if not only locally. In order to take care of the metric factor in the 
integration measure it is convenient to introduce an 
effective/external single particle geometrical potential 
$-\ln\sqrt{g(\rr)}/\tau$.

In the simulation we choose $\Delta_\theta$ and $\Delta_\varphi$ so to have
acceptance ratios as close as possible to $1/2$ in the acceptance/rejection 
rule for the random walk transition displacement moves of the Metropolis 
algorithm. The transition probability distribution function for the 
displacement move of the Metropolis algorithm will be uniform so it will 
drop out of the acceptance probability distribution function.

\section{The transition bridge move}
\label{app:bmove}

In order to take into account the particles permutations it is necessary
to construct two Brownian bridges between two different
\footnote{A bridge between the same particle can still be used to sample the
density matrix of distinguishable particles as can be done with the displacement
move of Appendix \ref{app:dmove}.}
randomly chosen particles in two randomly chosen many body beads to generate 
an exchange between the two particles. With
one bridge we connect particle $1$ on bead $R_i$ to particle $2$ on bead
$R_j$ and with the other we connect particle $2$ on bead $R_i$ to particle
$1$ on bead $R_j$ with $i<j$. This will produce an {\sl exchange} of particles
1 and 2.

The Brownian bridge between particle $1$ at $\rr_{1,i}$ and particle $2$ at 
$\rr_{2,j}$ is built through the following multislice transition move  
\cite{Ceperley1995},
\bq \label{eq:bmove1}
\xx_{{\rm new},i}&=&\xx_{1,i}\\ \label{eq:bmove2}
\xx_{{\rm new},k}&=&\xx_{{\rm new},k-1}+
\frac{\xx_{2,j}-\xx_{{\rm new},k-1}}{j-k+1}+
\xi~~~~~~k=i+1,\ldots,j
\eq
where $\xx=(\calx,\caly)$ is a two dimensional vector on a flat space   
and $\xi$ is a random number with a Gaussian probability distribution
\footnote{This can be generated with the Box-Muller algorithm 
\cite{Kalos-Whitlock} for example.}
with zero mean and variance $\sigma^2(j-k)/(j-k+1)$ where 
$\sigma^2=2\lambda\tau$ is the diagonal free particle variance. We will
then first perform a direct mapping $\theta\to\calx$ and $\varphi\to\caly$.
Then the bridge move of Eqs. (\ref{eq:bmove1})-(\ref{eq:bmove2}). And
finally we go back to the sphere with the inverse mapping 
$\calx\to\theta$ and $\caly\to\varphi$. 
We will then have a direct mapping of the many body system 
$R\to X=(\xx_1,\xx_2,\ldots,\xx_N)=(\{\xx_i\})$ and an inverse 
mapping $X\to R$ back on the sphere.

The Metropolis (rejection) method can sample any probability distribution
provided that the transition rule satisfies detailed balance and ergodicity.
The Metropolis algorithm is a particular way of ensuring that
the transition rule satisfies detailed balance. It does this by
splitting the {transition probability} into an ``a priori'' {{\sl sampling
distribution}} $T(s\to s')$ (which is a probability distribution that we
can sample) and an {{\sl acceptance probability}} $A(s\to s')$ where
$0\leq A\leq 1$.
\bq
P(s\to s')=T(s\to s')A(s\to s'),
\eq
In the generalized Metropolis procedure \cite{Kalos-Whitlock},
trial moves are accepted according to:
\bq \label{eq:acceptance}
A(s\to s')=\min[1,q(s\to s')],
\eq
where
\bq
q(s\to s')=\frac{\pi(s')T(s'\to s)}{\pi(s)T(s\to s')}.
\eq
where $\pi\propto e^{-\cals}$ is the action probability distribution in 
the $s=(\{R_k\},\calp)$ configurations space. The transition probability 
corresponding to the move of Eqs. (\ref{eq:bmove1})-(\ref{eq:bmove2}) is 
then given by
\bq
\frac{T(s_{\rm new}\to s_{\rm old})}{T(s_{\rm old}\to s_{\rm new})}
\propto\exp\left[
\sum_{k=i+1}^{j}(X_{{\rm new},k}-X_{{\rm new},k-1})^2/4\lambda\tau
-\sum_{k=i+1}^{j}(X_{{\rm old},k}-X_{{\rm old},k-1})^2/4\lambda\tau\right].
\eq

So we start from $R_{\rm old}$ map it to $X_{\rm old}$, on the direct mapping
flat space perform the Browninan bridge move, accept or reject the transition 
according to the probability (\ref{eq:acceptance}) to find $X_{\rm new}$, 
and finally inverse map it back to $R_{\rm new}$ on the sphere. Ergodicity 
in $\caly$ would be lost only if we are {\sl exactly} on a pole. But this is 
always prevented on a computer due to the finite arithmetic! Moreover since 
we have ergodicity everywhere in $\calx$ it will always be possible to escape
a pole by moving along $\calx$.

In order to produce an exchange of two particles $1$ and $2$ one needs a 
combination of two bridge transitions as described above.
Any permutation can be reached through a two particles exchange so the 
bridge transition move allows to sample the sum in Eq. (\ref{eq:dmid}).

\section{Braids sampling}
\label{app:cmove}

In order to sample the sum in Eq. (\ref{eq:dmidim}) one needs a move 
able to bring the path $R(t)$ from one homotopy class to another. And a way to 
understand to which homotopy class the path belongs after each move.
This will allow one to determine to which $\tilde{\rho}_\alpha$ he is 
contributing at each accepted transition move.

A path $R(t)$ made of closed distinct single particle paths,
$\rr(0)=\rr(\beta)$, on a sphere, can only have an even number $n$ of 
two particles crossings, i.e. of braids. Allowing for particles
exchanges $n$ can be any integer. 
If two different particles, say 1 and 2, have $m$ crossings 
between timeslices $k_i$ and $k_f$, an acceptance of the swap move described 
in Appendix \ref{app:bmove}, will necessarily result in a single crossing
if $m$ is even, and consequently $n\to n-m+1$, or to no crossing at all, 
if $m$ is odd, and consequently $n\to n-m$. 

We see then how we can use the swap move to jump from any homotopy class 
to any other. And within an homotopy class $\alpha$ we can then act with 
the displace move described in Appendix \ref{app:dmove} to sample 
$\tilde{\rho}_\alpha$.

In order to count the number of crossings between particles 1 and 2 between 
timeslices $k_i$ and $k_f$ it is necessary to count the number of times 
in which $\rr_{1,k}=\rr_{2,k}$ for $k\in[k_i,k_f]$. In order to take
into account of this crossing condition in the discretized imaginary 
time one can for example determine when both 
$r^\alpha_{1,k}-r^\alpha_{2,k}$ for $\alpha=1,2$ change sign or are zero, 
varying $k$. Counting the number $n$ of braids reached in the path $R(t)$ 
allows to asses to which path homotopy class one is contributing in the
density matrix. So at each move we need to count the number of crossings $n$
in the whole many body path $R_k$ from $k_i=0$ to $k_f=M$.


\section*{Author declarations}

\subsection*{Conflicts of interest}
None declared.

\subsection*{Data availability}
The data that support the findings of this study are available from the 
corresponding author upon reasonable request.

\subsection*{Funding}
None declared.

\bibliography{pimcsph}

\end{document}